\documentstyle[12pt]{article}

\author{M.I.Krivoruchenko\\
Institute for Theoretical and Experimental Physics,\\
B.Cheremushkinskaya, 25, 117259 Moscow, Russia\\
(e-mail: mikhail@vxitep.itep.ru)}
\title{A Constraint to the Parity-Conserving Parameter \\
of the Neutron-Antineutron Oscillations
}
\date{}
\begin{document}

\maketitle
\begin{abstract}
The phenomenology of neutrino oscillations is extended to the
neutron-anti-neutron oscillations predicted by the grand unified theories.
There are six parameters in general describing the one-flavor oscillations:
two real Majorana masses (of opposite signs), three Euler's angles, and a
phase multiplier $\chi $. The sum of the Majorana masses, treated as a small
parity-violating parameter $\epsilon $, induces the nuclear decays $%
(A,Z)\rightarrow (A-2,Z)$ and determines the oscillation rate of neutrons in
the vacuum. We derive a constraint to a small parity-conserving parameter $%
\epsilon ^{\prime }$ representing a combination of two Euler's angles. This
parameter gives no significant contribution to the neutron oscillation rate
in the vacuum, while contributes to the nuclear decay rate. The nuclear
stability implies $|\epsilon ^{\prime }|<1yr^{-1}$. Absence of the vacuum
neutron oscillations together with existence of the nuclear decays $%
(A,Z)\rightarrow (A-2,Z)$ would become simplest manifestation of the $%
\epsilon ^{\prime }\not =0$.
\end{abstract}

\vspace{5mm}

{\bf PACS:} 12.10.-g; 21.90.+f

\newpage

\setcounter{equation}{0}

Prediction$^1$ of the neutron-antineutron oscillations in the grand unified
theories (GUT's) stimulated searches for the neutron-antineutron
oscillations in the vacuum and in nuclei. The existing experiments constrain$%
^{1-10}$ the vacuum oscillation time to be $\tau =1/\mid \epsilon \mid >1$
yr.

The phenomenology of the neutron oscillations is identical to the
phenomenology of the one-flavor neutrino oscillations$^{11-14}$. The neutron
is described by a superposition of two real Majorana fields $\varphi _k$
with different masses $\mu _k$. The free neutron Lagrangian takes the form
\begin{equation}
\label{1}L_f=\frac 12\sum_{k=1,2}\bar \varphi _k(i\hat \nabla -\mu
_k)\varphi _k
\end{equation}
where $\bar \varphi _k=\varphi _k\gamma _0$. The fields $\varphi _1$ and $%
\varphi _2$ are expressible in terms of the complex neutron field $\psi $ as
follows
\begin{equation}
\label{2}\varphi _{kL}=a_k\psi _L+b_k\psi _L^{*}.
\end{equation}
We remind that in the Majorana representation $\gamma _\mu ^{*}=\tilde
\gamma ^\mu =-\gamma _\mu ,$ $\gamma _5^{*}=\tilde \gamma _5=-\gamma _5,\psi
_c=\psi ^{*}.$ The left components are given by $\psi _L=\frac 12(1-\gamma
_5)\psi $, etc. It follows then $\varphi _{kR}=(\varphi _{kL})^{*}$.

The kinetic term of the Lagrangian takes the standard form $L_{kin}=\bar
\psi i\hat \nabla \psi $ for
\begin{equation}
\label{3}\sum_{k=1,2}a_ka_k^{*}=\sum_{k=1,2}b_kb_k^{*}=1,%
\sum_{k=1,2}a_kb_k^{*}=0.
\end{equation}
The mass term of the Lagrangian ($L_f=L_{kin}+L_m$) becomes
\begin{equation}
\label{4}-L_m=m_D\bar \psi _R\psi _L+m_L\bar \psi _{cR}\psi _L+m_R\bar \psi
_{cL}\psi _R+m_D^{*}\bar \psi _R\psi _L+m_L^{*}\bar \psi _L\psi
_{cR}+m_R^{*}\bar \psi _R\psi _{cL}
\end{equation}
where
\begin{equation}
\label{5}m_D=\sum_{k=1,2}a_kb_k\mu _k,\ m_L=\sum_{k=1,2}a_k^2\mu _k,
\ m_R^{*}=\sum_{k=1,2}b_k^2\mu _k.
\end{equation}

The coefficients $a_k$ and $b_k$ can be interpreted as components of two
orthogonal spinors. The general expressions for these coefficients are given
by $a_k=D^{(1/2)}(\alpha ,\beta ,\gamma )_{k1}e^{i\chi }$ and $%
b_k=D^{(1/2)}(\alpha ,\beta ,\gamma )_{k2}e^{i\chi }$ . The spin-1/2
rotation matrix has the form
\begin{equation}
\label{6}D^{(1/2)}(\alpha ,\beta ,\gamma )=\left(
\begin{array}{cc}
\cos (\frac \beta 2)e^{i(\alpha +\gamma )/2} & \cos (\frac \beta
2)e^{i(-\alpha +\gamma )/2} \\
-\sin (\frac \beta 2)e^{i(\alpha -\gamma )/2} & \cos (\frac \beta
2)e^{i(-\alpha -\gamma )/2}
\end{array}
\right) ,
\end{equation}
the Euler's angles $\alpha ,\beta ,\gamma $ are defined as in Ref.15, Ch.58.
The phase $\chi $ is introduced to generate a two-parameter family $\beta
=\gamma =0$ of the orthogonal spinors polarized in the $z$-direction.

The masses $m_D$, $m_L$, and $m_R$ become
\begin{equation}
\label{7}
\begin{array}{c}
2m_De^{-i\chi }=\sin \beta (\mu _1e^{i\gamma }-\mu _2e^{-i\gamma }), \\
2m_Le^{-i(\alpha +\chi )}=\cos {}^2(\frac \beta 2)\mu _1e^{i\gamma }+\sin
{}^2(\frac \beta 2)\mu _2e^{-i\gamma }, \\
2m_L^{*}e^{i(\alpha -\chi )}=\sin {}^2(\frac \beta 2)\mu _1e^{i\gamma }+\cos
{}^2(\frac \beta 2)\mu _2e^{-i\gamma }.
\end{array}
\end{equation}
Here $\bar \psi _{cR}=((\psi _c)_R)^{+}\gamma _0$, etc. Given that complex
masses $m_D$, $m_L$, $m_R$ are known, one can find the diagonal masses $\mu
_k$, the Euler's angles $\alpha ,\beta ,\gamma $, and the phase $\chi $:
\begin{equation}
\label{8}
\begin{array}{c}
\tan \beta =m_D/(m_Le^{-i\alpha }-m_R^{*}e^{i\alpha }), \\
\mu _ke^{i(\pm \gamma +\chi )}=m_Le^{-i\alpha }+m_R^{*}e^{i\alpha }\pm \sqrt{%
(m_Le^{-i\alpha }-m_R^{*}e^{i\alpha })^2+m_D^2}.
\end{array}
\end{equation}
The sign $(+)$ stands for $k=1$. The values $\alpha ,\beta ,\gamma ,$ and $%
\chi $ are fixed from a requirement that $\tan \beta $ and $\mu _k$ be real.
The mass term of the Lagrangian contains tree complex masses $m_D$, $m_L$,
and $m_R$ and therefore six independent parameters. These parameters are
described in terms of the two diagonal masses $\mu _k$, three Euler's angles
$\alpha ,\beta ,\gamma ,$ and one phase $\chi .$ The phase transformation $%
\psi \rightarrow e^{-i\alpha /2}\psi $ can be made to set $\alpha $ equal to
zero. The chiral transformation $\psi \rightarrow e^{i\gamma _5\chi /2}\psi $
removes the phase $\chi $ from Eqs.(7) and (8).

The Lagrangian (2) can finally be written in the form
\begin{equation}
\label{9}L_f=\bar \psi (i\hat \nabla -m)\psi -\frac 12\Delta \bar \psi
i\gamma _5\psi -\frac 12\epsilon \bar \psi _c\psi -\frac 12\epsilon ^{*}\bar
\psi \psi _c-\frac 12\epsilon ^{\prime }\bar \psi _ci\gamma _5\psi -\frac
12\epsilon ^{\prime *}\bar \psi i\gamma _5\psi _c
\end{equation}
where $m=(m_D+m_D^{*})/2,$ $\Delta =i(m_D-m_D^{*}),$ $\epsilon =m_L+m_R,$
and $\epsilon ^{\prime }=i(m_L-m_R)$.

The choice of signs of the $\mu _k$ is a matter of convention. We assume $%
\mu _1\approx -\mu _2>0$. It follows then that we are working in a region of
small values $\mu _1+\mu _2$, $\pi /2-\beta ,$ $\gamma $ and $\chi $. To the
lowest order in these parameters, we get
\begin{equation}
\label{10}
\begin{array}{c}
m=\frac 12\sin \beta (\mu _1\cos (\gamma +\chi )-\mu _2\cos (\gamma -\chi ))
\\
\approx \frac 12(\mu _1-\mu _2), \\
\Delta =-\sin \beta (\mu _1\cos (\gamma +\chi )+\mu _2\cos (\gamma -\chi ))
\\
\approx -(\mu _1-\mu _2)\chi , \\
\epsilon =e^{i\alpha }\frac 12(e^{i\gamma }\cos {}^2(\frac \beta 2)(\mu
_1e^{i\chi }+\mu _2e^{-i\chi })+e^{-i\gamma }\sin {}^2(\frac \beta 2)(\mu
_1e^{-i\chi }+\mu _2e^{i\chi })) \\
\approx e^{i\alpha }\frac 12(\mu _1+\mu _2), \\
\epsilon ^{\prime }=ie^{i\alpha }\frac 12(e^{i\gamma }\cos {}^2(\frac \beta
2)(\mu _1e^{i\chi }-\mu _2e^{-i\chi })-e^{-i\gamma }\sin {}^2(\frac \beta
2)(\mu _1e^{-i\chi }-\mu _2e^{i\chi })) \\
\approx ie^{i\alpha }\frac 12(\mu _1-\mu _2)(\pi /2-\beta +i\gamma ).
\end{array}
\end{equation}

In the external electromagnetic field, the neutron Lagrangian acquires a
term $L_{int}=$ $\bar \psi \frac 12\mu _n\sigma _{\mu \nu }F_{\mu \nu }\psi $%
, with $\mu _n$ being the neutron magnetic moment. The chiral transformation
$\psi \rightarrow e^{i\gamma _5\chi /2}\psi $ with $2m\chi =-\Delta $
removes the second term from Eq.(9), generating an interaction $%
L_{int}^{\prime }=\bar \psi \frac 12d\sigma _{\mu \nu }\tilde F_{\mu \nu
}\psi $ with $\tilde F_{\mu \nu }=\frac 12\epsilon _{\mu \nu \tau \upsilon
}F_{\tau \upsilon }$. The parameter $\Delta $ is related to the neutron
electric dipole moment $d=\mu _n\frac \Delta {2m}.\ $Limits to the dipole
moment $d$ are, however, many orders of magnitude lower then a value
expected from the GUT's.

The vacuum oscillation rate to the lowest order is determined solely by
parameter $\epsilon $. The neutron oscillation time in the vacuum is $\tau
=1/|\epsilon |$.

The nuclear stability poses, however, comparable constraints to the $%
\epsilon $ and $\epsilon ^{\prime }$, since the nuclear decays are equally
well induced by terms of the Lagrangian (9), proportional to the values $%
\epsilon $ and $\epsilon ^{\prime }$. There is no interference between the
corresponding terms, since they induce the opposite-parity transitions. The
nuclear width can therefore be represented by a sum $\Gamma =\Gamma
_{-}+\Gamma _{+}.$ The value $\Gamma _{-}\propto |\epsilon |^2$ describes
the parity-violating transitions, whereas the value $\Gamma _{+}\propto
|\epsilon ^{\prime }|^2$ describes the parity-conserving transitions. The
first kind of the transitions is well known$^{1-10}$.

Shown in Fig.1 is a typical diagram contributing the nuclear decay $%
(A,Z)\rightarrow (A-2,Z)$ accompanied by emission of $\pi $-mesons. The
parity-violating transitions differ from the parity-conserving transitions
only by form of the vertex describing the neutron-antineutron annihilation.
In the first case the vertex is proportional to the $\epsilon $, whereas in
the second case it is proportional to the $\epsilon ^{\prime }i\gamma _5$.
The matrix $i\gamma _5$ mixes the upper and lower bispinor components. The
energy release in the nuclear decay is high (about two nucleon masses), so
such a mixing cannot produce a suppression $(v_F/c)^2$ with $v_F$ being the
Fermi velocity of nucleons, despite the decaying nucleus represents a
nonrelativistic system. We get therefore
\begin{equation}
\label{11}\Gamma _{+}/|\epsilon ^{\prime }|^2\sim \Gamma _{-}/|\epsilon |^2.
\end{equation}
Using estimates$^{1-8}$ for the parity-violating nuclear width $\Gamma _{-}$
and results for testing the nuclear stability$^{10}$ we derive a constraint
\begin{equation}
\label{12}|\epsilon ^{\prime }|<1yr^{-1}.
\end{equation}

An observation of the vacuum neutron oscillations would give an evidence for
the $\epsilon \not =0$. At the same time, an observation of the nuclear
decays $(A,Z)\rightarrow (A-2,Z)$ together with zero or low vacuum
oscillation rate of neutrons would give a direct experimental evidence for
the $\epsilon ^{\prime }\not =0$.

The author is indebted to B.V.Martemyanov and M.G.Schepkin for useful
discussions. This work is supported by the Russian Foundation for
Fundamental Researches under the Grant No. 94-02-03068 and the INTAS\ Grant
No. 93-0079.

\end{document}